\documentclass[aps,prl,twocolumn,superscriptaddress]{revtex4-2}

\usepackage{graphicx}
\usepackage{dcolumn}
\usepackage{bm}
\usepackage{ulem}
\usepackage{xcolor}
\usepackage{amsmath}

\begin{document}


\title{Spatiotemporal statistics of  the dissipation rate at the boundary of  a turbulent flow using Diffusing-Wave Spectroscopy}

\author{Enzo Francisco}
\author{Julien Lambret}
\author{Sébastien Aumaître}
\email{sebastien.aumaitre@cea.fr}
\affiliation{
Service de Physique de l’État Condensé, Université Paris-Saclay, CNRS, CEA \\
91191 Gif‑sur‑Yvette Cedex, France
}%

\date{\today}

\begin{abstract}
{We use Diffusing Wave Spectroscopy (DWS) to perform the first direct space- and time-resolved measurement of the dissipation rate~$\epsilon$ at the boundary of a turbulent flow. We have shown in a previous publication that this technique provides maps of the dissipation rate of Newtonian fluids~\cite{Francisco}. 
Here, we apply the technique at the boundary of a turbulent flow generated in a square box by an impeller stirring the fluids. Although the measurement is made on a small region near the boundary, we show that the dissipation remains proportional to the injected power and follows the turbulent scaling $\epsilon \propto \mathrm{Re}^3$, with Re being the Reynolds number ranging from $1.5 \times 10^4$ to $6 \times 10^5$.
With this flow, there is no need for logarithmic corrections to reproduce the dissipation near the flat boundary. In addition, our setup allows us to measure the spatio-temporal fluctuations of the dissipation near the boundary. These fluctuations are quite large (the relative fluctuations are about 50\%) and are well described by a log-normal distribution, as expected for the dissipation rate in the bulk of homogeneous and isotropic turbulence (HIT) but Power Density Spectra (PDS) do not correspond to those expected for HIT \cite{Li07,Graham16,K62}.
 }\\
\end{abstract}

\maketitle

{\bf Introduction:}
Turbulence is a complex phenomenon in which dissipation is a key quantity. Indeed, the dissipation rate determines the energy consumption of vehicles and pressure losses in pipes \cite{Ahmed84, Lawn71}. At the geophysical scale, it plays a role in the energy balance and transport efficiency in climate models \cite{Fiedler00,Asaro11}. At a fundamental level, the dissipative structures might drive turbulence intermittency \cite{Yaglom,Frish, K62}. Nevertheless, it is very difficult to measure dissipation experimentally \cite{Guichao2021}. Indeed, the dissipation rate in a Newtonian fluid flow with a velocity field $\mathbf{u}$ is given by  
\begin{equation}
\epsilon=\sum_{i,j}\frac{\nu}{2} \left(\partial_j u_i +\partial_i u_j \right)^2,
\label{epsilon}
\end{equation} 
where $\nu$ is the kinematic viscosity and where $\{i,j\}$ stand for Cartesian coordinates $\{x,y,z\}$. Hence, it requires knowledge of the norm of the strain-rate tensor $\partial_j u_i$, i.e., the measurement of the spatial derivatives in all directions of the entire vector field $u_i$. Nevertheless, as the velocity gradient is a key quantity in fluid mechanics, many attempts have been made to measure it, but it is not an easy task. With hot-wire techniques, Taylor's hypothesis of frozen turbulence must be assumed, and probes made up of many wires become perturbative \cite{Tsinober1992}. Particle image velocimetry has also been used successfully, either by focusing on a point measurement \cite{LathropNature}, or by resolving the dissipation rate in a plane using the complex technique of dual-plane stereo particle image velocimetry \cite{Mullin}. In both cases, the maximum accuracy of the gradients accessible by the measure is linked to the size of the grid used for the interrogation window. These techniques are therefore only applicable to a limited range of moderate Reynolds numbers. Direct Numerical Simulations also fail to estimate such a quantity over long times because of the high numerical cost, and thus may underestimate rare events \cite{Guichao2021,Schumacher}. These difficulties are amplified in many cases where we are interested by the dissipation and the shear rate near a surface. For instance one can mention the drag of an immersed object, the erosion efficiency of a geophysical flow, or canopy dynamics \cite{Dosanjh85,Hogg97,Finnigan00}. In smooth channel or pipe flows, where turbulence develops from the wall, one can recover the wall friction by balancing it with the pressure drop along the channel. In that case, a logarithmic correction is predicted and measured experimentally for the wall friction \cite{Yaglom,Landau}. In our device, where the power is injected by an impeller with blades, there is no prediction for the friction at the boundary.

Here, we present an experimental technique allowing us to measure directly $\epsilon$. The method is based on Diffusing Wave Spectroscopy (DWS) applied on a turbid fluid. It was developed in the 80s on very simple flows \cite{Pine88,Pine91,Bicout93,Bicout94}. The coherent light scattered by a turbid fluid in motion was measured with a Photomultiplier Tube (PMT), which is very sensitive and fast but did not allow spatio-temporal measurements. Then it was applied to complex fluids rheology, granular flows, or solid deformations (see  \cite{Durian91,Horn93,Menon97,Mason97,Palmer99,CohenAdda01,Crassous14} for a few examples).

Recently, we have shown that the technical improvements of high-speed imaging enable a quantitative spatio-temporal measurement of the dissipation at the boundary of the well-known Taylor-Couette flow \cite{Francisco}. In this letter, we show that it can be applied successfully to a fully turbulent flow. It opens a new window on turbulent phenomena by providing access to the first direct measurements of spatio-temporal maps of dissipation at the boundary of a turbulent flow, with a degree of precision and versatility not possible with other methods. In our setup, where a turbulent flow is generated in a square container by an impeller, we resolve surprisingly large fluctuations of the dissipative structures up to Reynolds number $Re=6\times10^5$.

This letter is organized as follows. We first briefly recall the principle of the method and its limitations. Then we describe our experimental setup and the measurement protocol. The global scaling of the estimated dissipation as a function of $Re$ is presented. We finally focus on the properties of the fluctuations of the dissipation. We determine the Probability Density Function (PDF) and Power Density Spectrum (PDS) of the measured dissipation and we compare them to the Direct Numerical Simulations of the John Hopkins Database \cite{Li07,Graham16}.
 
{\bf Theoretical background:}
We detail the DWS applied to turbid fluid flows in a previous publication \cite{Francisco} where we applied the DWS on the well-known Taylor-Couette flow. It allows us to demonstrate that the techniques resolved spatio-temporally the dissipation in quantitive manner. We recall here only the principle and requirements. A coherent light illuminates a moving fluid seeded with small particles that multi-scatter the incident beams. The beams escaping the fluid interfere in a speckle pattern. Due to the motion of the scatterers in the fluid, the speckles sparkle. If the radius of the particles, $r$, and the wavelength of the light, $\lambda$, are small compared to the transport mean free path of the light $l^*$, and if the characteristic length of the gradients is  large compared to the optical path through the fluid, then we can write the self-correlation function of the back-scattered intensity, $I(t)$:
\begin{equation}
\begin{aligned}
g_2(\tau)&=\frac{\langle I(t+\tau) I(t) \rangle}{\langle I(t+\tau) \rangle\langle I(t) \rangle}\\
&=\beta \exp(-\gamma\sqrt{6(\tau/\tau_o+\tau^2/\tau_V^2)})+1
\label{g2_2}
\end{aligned}
\end{equation}

where $\beta$ is the intensity contrast ($\langle I^2 \rangle/\langle I\rangle^2-1$) and $\gamma$ is a parameter depending only on the optical geometry and boundary conditions. $\tau_o$ and $\tau_V$ are related to the scatterers' displacement. $\tau_o=1/(Dk^2)$ is due to Brownian motion. $D$ is the diffusion coefficient of the scattering particles. It is known through the Stokes--Einstein Formula, giving $D=k_BT/(6\pi r \mu)$, with $k_B$ the Boltzmann constant and $\mu$ the dynamic viscosity. $k$ is the wavenumber. $\tau_V$ is due to the fluid motion and can be directly related to $\epsilon$ by:

\begin{equation}
\tau_V=\sqrt{30}/\left(l^* k \sqrt{\langle \epsilon\rangle /\nu}\right)
\label{tauV}
\end{equation}
where $\langle\cdot\rangle$ stands for a spatial average over the volume explored by the scattered beams in the fluid. It can occur in two ways: (i) If $I$ is measured in the far field with a Photo-Multiplier Tube (PMT), one probes the volume given by the penetration depth, a few $l^*$ in back-scattering configuration, times the surface formed at the cell surface by all the collected beams. It depends on the numerical aperture of the single-mode fiber mounted on the PMT and its distance to the cell. (ii) We can also use a high-speed camera focused on the surface. In that case, the outgoing beams have explored most probably a volume of order $l^{*3}$ around their escaping point. We adjust the optics of the camera such that the pixels of the camera on the cell are larger than $l^*$. In this case, each pixel gives local information averaged over its surface times the penetration depth. As we impose $\lambda\ll l^*$ to be in the diffusive approximation, $\tau_V\propto \lambda/l^*$ is much shorter than $1/ \sqrt{\langle \epsilon\rangle/\nu}$. Hence, we can capture the time evolution of $\langle \epsilon\rangle$ if we converge $g_2(\tau)$ over a time smaller than the flow evolution.

\begin{figure}
\begin{center}
\includegraphics[width=8cm]{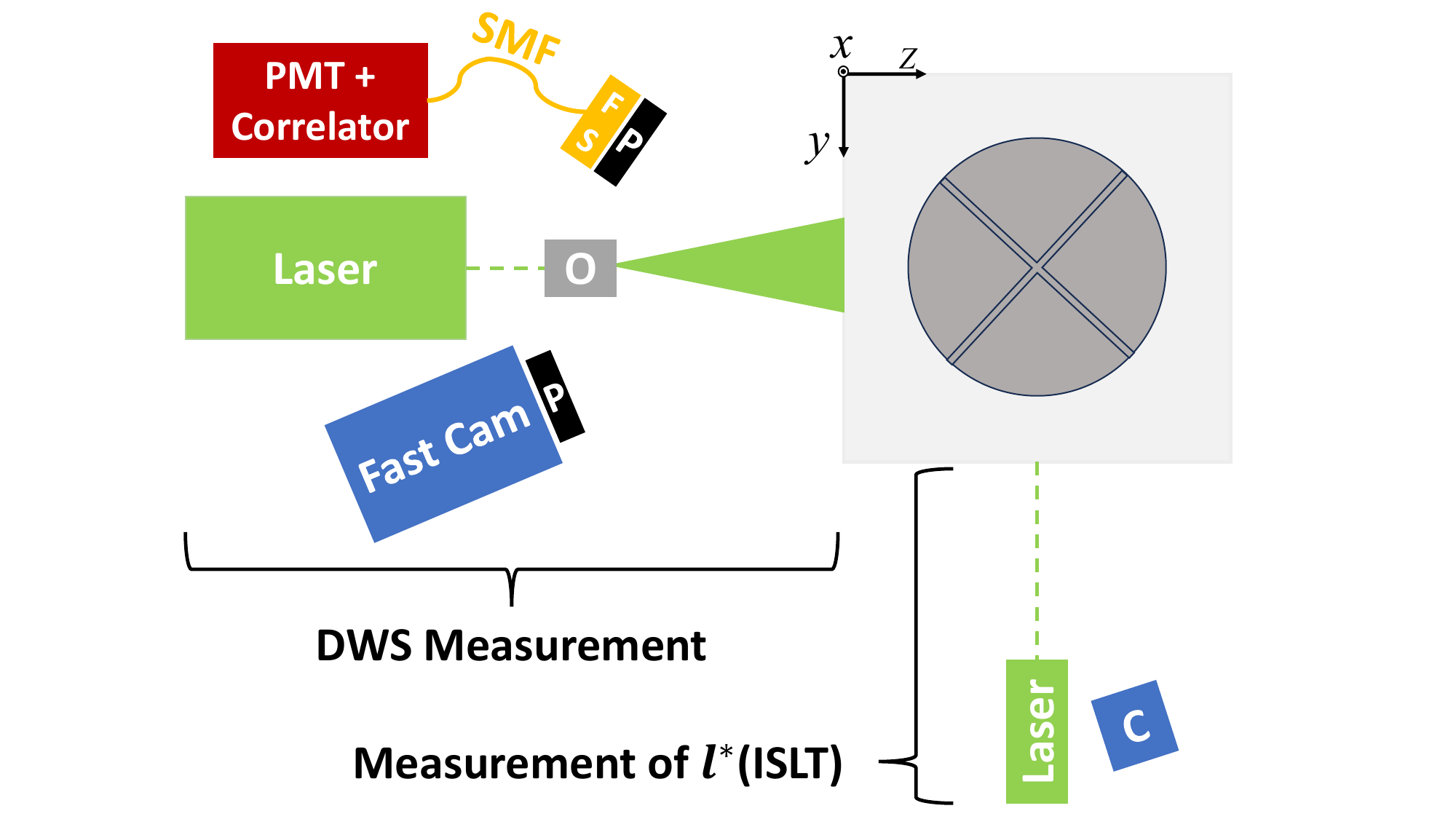}
\end{center}
\begin{center}
\includegraphics[width=8cm]{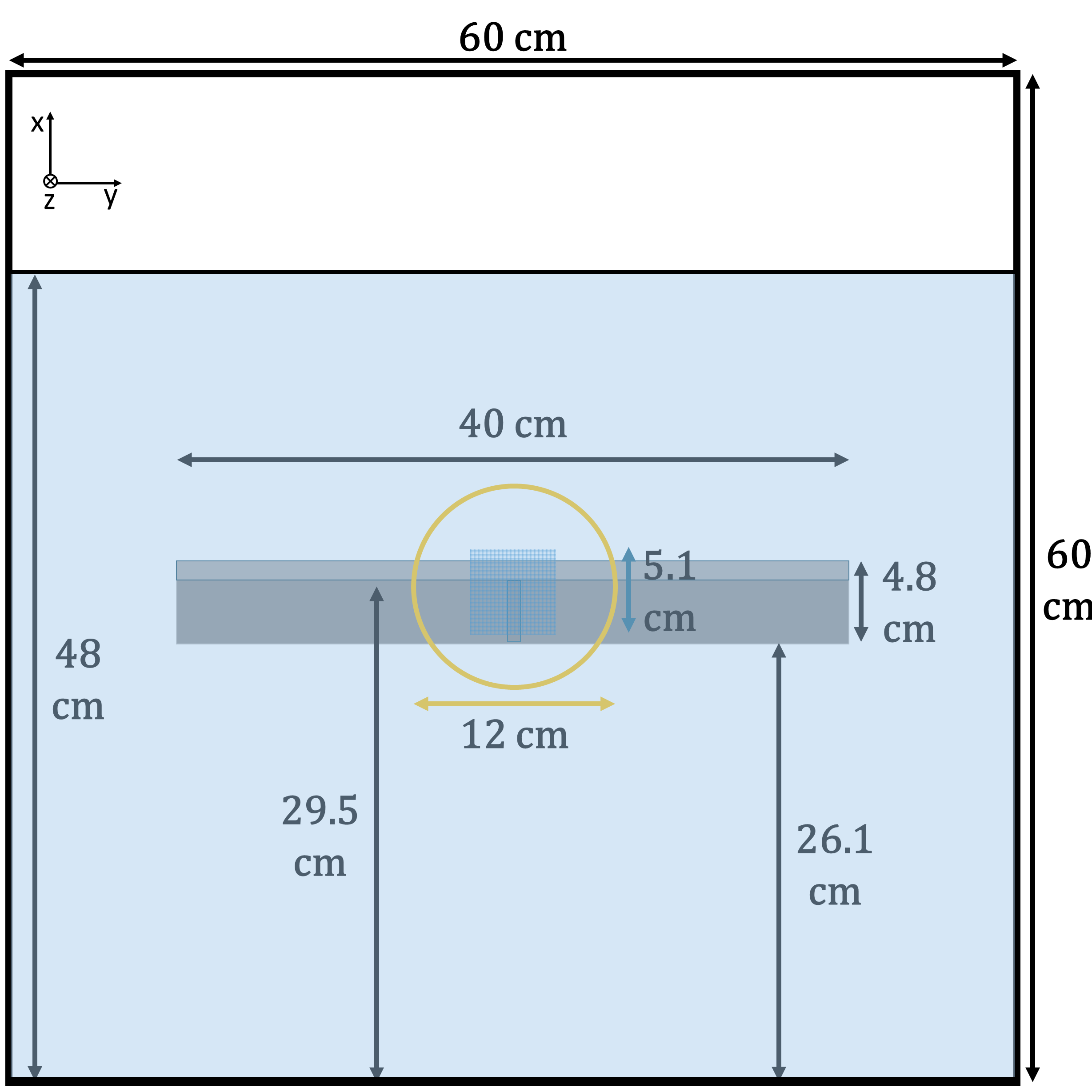}

\caption{{\bf Top} sketch of the experimental setup viewed from the top. The DWS involves a 2-Watt Neodymium-YAG laser illuminating a large area of the cell through a microscope lens. The diffused light is collected either in the far field by a Photo Multiplier Tube (PMT) through a cross-polarizer or by an ultra-fast camera. We continuously control the value of $l^*$ with the spatially resolved reflectance technique involving another Neodymium-YAG laser and a CCD camera. {\bf Bottom} Side view of the experimental cell. The  gray rectangle shows the impeller position with the disk (light gray) and the four blades (dark gray). The blue square represents the area measured with the high-speed camera. The orange circle delimits the light collected by the monomodal fiber.
}
\label{Setup}
\end{center}
\end{figure}

{\bf Order of magnitude and experimental setup:}
The time necessary to converge the correlation functions of the light collected with the high-speed camera is indeed the main constraint of the method. Typically, it cannot be less than 0.06 s with our equipment. Therefore, to capture the dynamics of the turbulent flow, one has to design a setup with a large $Re$ but a low characteristic time. This can be reached with a relatively large device. With that in mind, we built the experimental setup drawn in figure \ref{Setup}. The flow is generated in a square aquarium of size $L= 60$ cm by an impeller of radius $R=20$ cm. This impeller consists of a 1.2 cm thick disc and four 3.6 cm high blades to stir the fluid. It is rotating around the x-axis, generating a mean toroidal flow in the y-direction (see figure \ref{Setup}) (the impeller also generates a upward poloidal pumping flow . It is driven  by a step motor mounted on a gear reducer in order to reach the desired range of rotation speeds $\Omega$ going from 0.02 to 0.8 rotations per second. This imposes a macroscopic time scale of order $1/(4\Omega)$ ranging from 0.3 to 12.5 s.

In terms of dimensionless parameters, one can define the Reynolds number $Re=2\pi UL/\nu$ with the characteristic velocity $U=R\Omega$ and where the kinematic viscosity of the suspension is equal to that of pure water: $\nu\sim 1\times 10^{-6}$ m$^2$.s$^{-1}$. With this definition, we get $Re\in [1.5\times 10^4 ; 6\times 10^5]$ ($Re\in [2.4\times 10^4 ; 3.8\times 10^5]$ with the high-speed camera). A torquemeter is implemented between the motor and the impeller to monitor the rotation speed $\Omega$ and measure the torque $C$. We then compute the full injected power $P= C\cdot\Omega$.

The cell is filled with 170 l of deionized water seeded with {\bf 1.7 kg of TiO$_2$ particles} (corresponding to 1\% in mass) of mean hydrodynamic radius $r=200$ nm. This size guarantees a passive tracer behavior of the scatterers with a Stokes number
$S_k=\frac{\rho_{s}}{18\rho_w}\frac{2r}{L}Re\leq 0.1$ with $\rho_{s}$ and $\rho_{w}$ the densities of the TiO$_2$ particles and water respectively. The particles smallness also prevents the settling of the particles despite their density $\rho_s>\rho_w$. The settling velocity, $U_s$, obtained by balancing gravity force, buoyancy, and viscous friction:  $4\pi/3(\rho_{s}-\rho_w) r^3 g=6\pi \rho_w \nu rU_s$, is less than $10^{-7}  \mu$m/s !). It implies a settling of 0.01 $\mu$m during a day (our experimental run never exceeded 2 days). The deionized water is necessary to ensure the stability of the suspension \cite{Przadka}. Such a suspension gives $l^*=80~\mu$m \cite{Francisco}.

The smallest scale expected in the bulk of a homogeneous and isotropic turbulent (HIT) flow is the Kolmogorov scale $\eta_K=L/Re^{3/4}$. It ranges from 440 $\mu$m to 30 $\mu$m in our experiment. Therefore, it crosses $l^*=80~\mu$m and we may expect that our measurement fails for $l^*\leq \eta_K$. Nevertheless, $\eta_K$ might not be the relevant length scale at the boundary of a turbulent flow. The typical size of the velocity fluctuations can be given by the Taylor microscale: $\Lambda_T/L=1/\sqrt{Re}$. One always has $\Lambda_T > l^*$. Another pertinent length scale in wall-bounded turbulence is the viscous sublayer. The transition between the viscous sublayer and the turbulent boundary layer is smooth and somewhat arbitrary. Following \cite{Yaglom}, we take $\delta_\nu/L\sim 25/Re$. We can expect $\delta_\nu\sim 1$ mm to $12.5~\mu$m. 

 The statistical weight of the optical path in the layer $\delta_\nu$ varies between $93.6\%$ and $2.9\%$ according to diffusion theory \cite{Sheng}. Hence we may expect a change of behavior in our range of $Re$. All these length scales are resumed in Table.1.
\begin{table*}[htbp]
\caption{\label{tab:table1} Characteristic length scale in the experiment}
\begin{ruledtabular}
\begin{tabular}{|c|c|c|c|c|c|c|c|}
$\lambda$&part. rad. r&$l^*$&Cell Size $L$& Imp. rad. $R$& Taylors $\Lambda_T$& Kolmogorov $\eta_K$& viscous sublayer $\delta_\nu$\\
\hline
0.532 $\mu$m&0.2$\mu$m&80$\mu$m& 60 cm& 20 cm&5-0.8 mm&440-30$\mu$m&1mm-25$\mu$m\\
\end{tabular}
\end{ruledtabular}
\end{table*}

The optical arrangement for the DWS is similar to that used in \cite{Francisco}. The cell is illuminated by a two-Watt Neodymium-YAG laser ($\lambda=532$ nm). A microscope lens ($\times$20) enlarges the beam illuminating the cell. We use both a single-mode optical fiber attached to a PMT and a correlator, and a high-speed camera to receive the diffused light selected with a polarizer perpendicular to the incident light. The fiber probes the average dissipation on a surface of 12 cm diameter, whereas the camera resolves the fluctuations of $\epsilon$ in a square of 128$\times$128 pixels corresponding to 5.1$\times$5.1 cm$^2$, i.e., a camera pixel corresponds to nearly 400 $\mu$m on the interface. The depth explored by the light in the fluid is less than $5l^*\sim 400~\mu$m. We fix the correlator sampling time to 1.28 $\mu$s. The frame rate of the camera is 430,000 images per second, but 25,000 images are necessary to converge the correlation function (\ref{g2_2}). The temporal resolution is then 0.06 s. It remains much smaller than $1/4\Omega$. One can notice that the requirements $r,\lambda\ll l^*$ are easily fulfilled.

In order to measure $l^*$ \textit{in situ}, we implement spatially resolved reflectance techniques with another Neodymium-YAG laser (250 mW, $\lambda=532$ nm) illuminating another face of the cell at the height of the impeller. We measure with a CCD camera the radial decay of intensity of the diffused light spot. It follows a law ($\propto (r/l^*)^{-3}$) from which we can deduce $l^*$ (see \cite{Kienle}). It appears that $l^*$ does not evolve during the 48 hours of our experimental campaigns, if we apply continuous stirring. We conducted the experimental runs as follows. We first compute $g_2(\tau)$ with the fluid at rest. Knowing $\tau_o$ with the Stokes-Einstein formula, we deduce $\gamma$. Then we run the impeller at a prescribed velocity (we usually alternate high and low stirring to keep the scatterers well mixed) and we determine the remaining unknown $\tau_V$ from equation (\ref{tauV}). Most of the measurements are relatively short (about 2 s because of the limited RAM available in the high-speed camera) except for one long measurement of 90 s (made of 45 separated movies of 2 s) at Re=$2\times10^5$. We end with a measurement at rest to double-check the value of $\gamma$.

\begin{figure}
\begin{center}
\includegraphics[width=8.7cm]{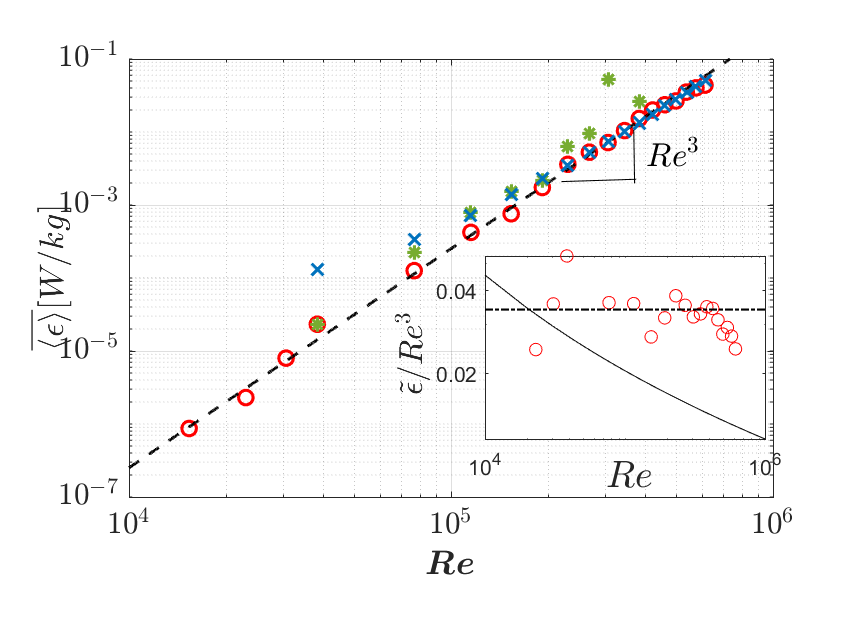}
\vspace{-0.5cm}
\caption{Scaling of the mean injected power per unit mass $\overline{P}$ (blue crosses) and the mean power dissipated per unit mass near the boundary. $\overline{\langle \epsilon \rangle}$ measured with the PMT (red circles) is averaged in time and over a disk of 12 cm diameter. For the fast camera (green asterisks), the measurement is averaged over all the pixels (corresponding to a square of 5.1 cm). $\overline{P}$ is multiplied by a factor 2.4 to align with the dissipated power, highlighting an excess of dissipation at our measurement spot. The inset compares the dimensionless dissipation $\tilde{\epsilon}=\overline{\langle \epsilon \rangle}(L/\nu)$ compensated by $Re^3$ (red circles) with the theoretical logarithmic correction (black line) or without (dot-dashed line).}
\label{Global}
\end{center}
\end{figure}

{\bf Results:}
In principle, our technique is more versatile than previous methods used to measure the dissipation. The accuracy  of the strain rate tensor estimation does not depend on the camera resolution. This allows us to explore a large range of Reynold number with the same experimental configuration. Therefore, we first study the global behavior of the dissipation measured either with the PMT or with the camera as the function $Re$ and compare it with the injected power. The main panel of figure \ref{Global} shows the dissipation $\overline{\langle\epsilon\rangle}$, where $\langle~\cdot~\rangle$ denotes a space average (over a volume of depth a few hundred $\mu$m times a surface area of 25 cm$^2$ with the camera or 113 cm$^2$ with the fiber) and where $\overline{~\cdot~}$  denotes a time average over the 2 s measurements. It also shows the mean injected power $\overline{P}$. Except at low $Re$, where the torque measurement is affected by the friction of the gears, $\overline{\langle\epsilon\rangle}$ and $\overline{P}$ are proportional, even though we probe only a small volume near the cell surface. Consequently, the dissipated power measured near the boundary follows the same scaling with Reynolds number as the injected power.  The zero law of turbulence imposes this scaling: $\overline{\langle \epsilon\rangle}\propto\overline{P}\propto (\nu^3/L^4)\cdot Re^3$. However, since we are considering footprint of the dissipation at the boundary of a turbulent flow generated by an impeller, we can expect some boundary layer signatures. Our measurements show that, unlike flows in channels and pipes, it is not necessary to apply logarithmic corrections to adjust the dissipation. The dissipation induced by the boundary layer is erased by the dissipative structure generated in the bulk (see the box in Figure \ref{Global}). 
Nevertheless, the dissipation measured per unit mass is about 2.4 times greater than the average power injected per unit mass. This is not surprising. Indeed, the flow generated by a single impeller in a square box is quite inhomogeneous and we do not expect a homogeneous spreading of the dissipative structures.  In fact, such flows exhibit excess dissipation at the cell boundary near the impeller \cite{Faller21}.

\begin{figure}
\begin{center}
\includegraphics[width=8.7cm]{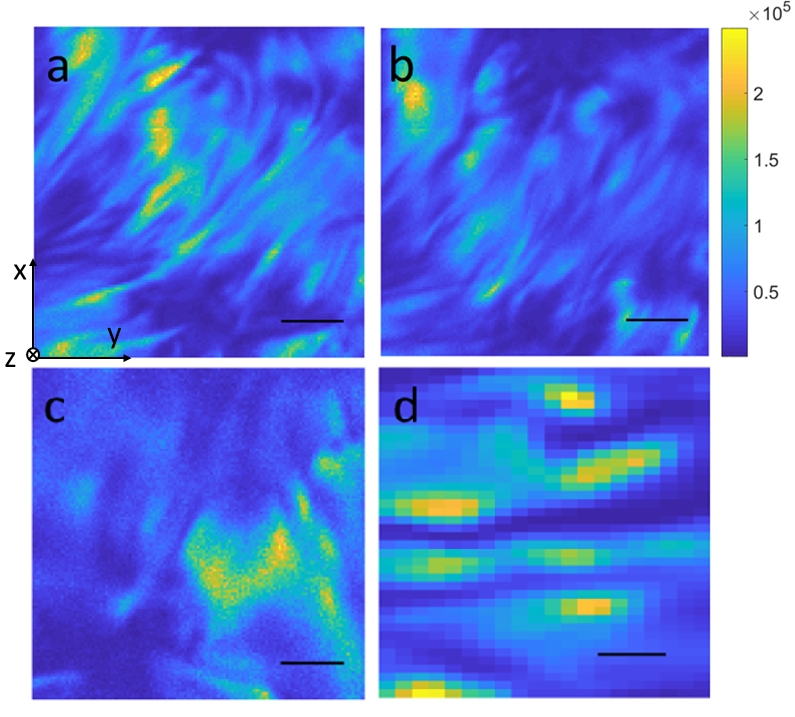}

\caption{Snapshots of dissipation maps in a window of $5.1\times 5.1$ cm$^2$ as shown by the blue square in figure \ref{Setup}. {\bf a} and {\bf b} show two successive maps from the experiment at $Re=3.2\times10^5$. The impeller rotates around the x-axis, generating a mean flow in the y-direction. {\bf c} shows a map from the experiment at $Re=1\times10^5$ and {\bf d} shows a map from DNS at $Re=3\times10^4$ (horizontal streamwise direction). The black line represents 1 cm.}
\label{Snapshots}
\end{center}
\end{figure}

Using the high-speed camera, we can resolve the spatial fluctuations of $\epsilon$ within the  5.1$\times$5.1 cm$^2$. Figures \ref{Snapshots}-a and \ref{Snapshots}-b show dissipation maps at two successive times for $Re=3\times10^5$. They reveal very large fluctuations, more than 50\%, within the probed volume. Although statistical convergence is not fully satisfied with the 2s movies, the data show that the standard deviation of the dissipation rate, $\sigma_\epsilon$, follows the same low: $\sigma_\epsilon\propto (\nu^3/L^4)\cdot Re^3$. Consequently, the relative fluctuations $\sigma_\epsilon/\langle \epsilon \rangle $ are independent of $Re$. The dissipative structures are significantly larger than our pixel size and are advected by the large-scale flow, which stretches them. In our device, driven by a single impeller with blades, we expect an average flow mostly toroidal with an upwelling poloidal component (see movies in the supplemental material). The correlator allows high-frequency measurements but the surface on which dissipation is averaged is important : a disk of 12 cm of diameter. Fluctuations are reduced by the spatial averaging although they remain of order of 30\%. Surprisingly, those fluctuations do not exhibit any frequencies connected to the passage of the blades (see the temporal trace in the supplemental materials). 

\begin{figure}
\begin{center}
\includegraphics[width=8.7cm]{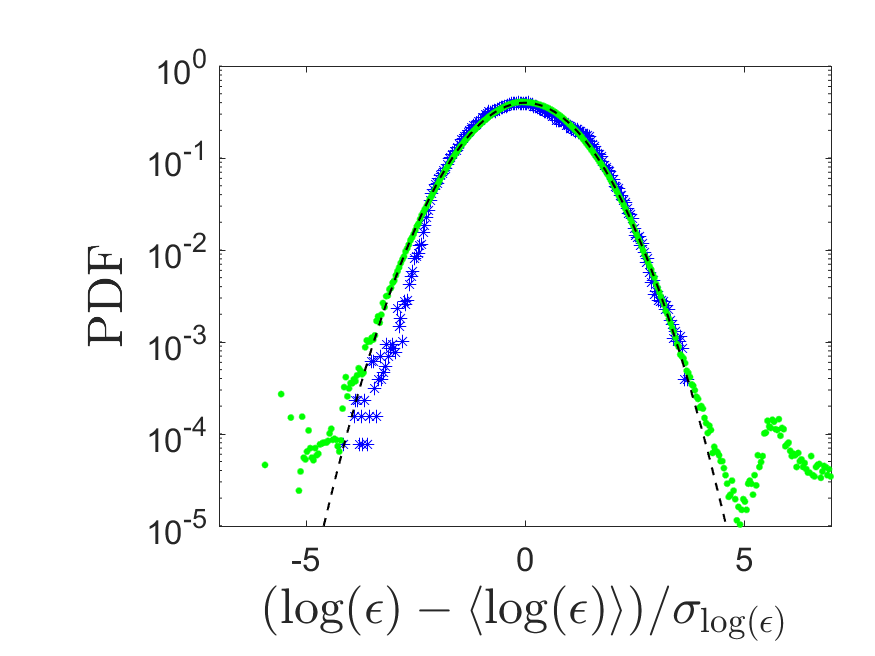}

\caption{Probability Density Function of the centered and normalized logarithm of dissipation fluctuations, $\left(\log(\epsilon)-\langle\log(\epsilon)\rangle\right)/\sigma_{\log(\epsilon)}$. Green dots correspond to the long measurement at $Re=3.2\times10^5$, blue dots correspond to DNS data from a channel flow. The dashed line represents the centered and normalized log-normal distribution.}
\label{PDF}
\end{center}
\end{figure}

A long measurement of 90 s (composed of 45 movies of 2 s each) was performed to establish the probability density function (PDF) of these fluctuations. Figure \ref{PDF} shows the resulting PDF of $\left(\log(\epsilon)-\langle\log(\epsilon)\rangle\right)/\sigma_{\log(\epsilon)}$ on a logarithmic y-axis. It is clear that the PDF of the dissipation fluctuations is compatible with a log-normal distribution. Such a distribution has been proposed to describe dissipation fluctuations in homogeneous isotropic turbulence (HIT). However, this description is based on the cascading processes specific to HIT. These processes might not be dominant in turbulent boundary layers, where complex hierarchical structures are known to exist \cite{Robinson91}. Furthermore, the refined Kolmogorov theory \cite{K62} for the HIT, prescribes fluctuations of the dissipation  such that:
\begin{equation}
\langle \epsilon\left(\mathbf{x}\right)\epsilon\left(\mathbf{x+r}\right)\rangle\propto \langle \epsilon^2 \rangle (L/r)^\mu
\end{equation}

imposing a Power spectum of $\epsilon$ in $k^{-(1-\mu)}$ \cite{Sreenivasan,Nelkin} where $\mu$ is the intermittent coefficient estimated between 0.2 and 0.3. This power law of the spectrum in the inertial range is in fairly good agreement with the estimate of the dissipation rate spectrum in atmopheric boundary layers \cite{Sreenivasan} and DNS of Taylor--Green vorticies \cite{Brachet}. 

\begin{figure}
\begin{center}
\includegraphics[width=8.7cm]{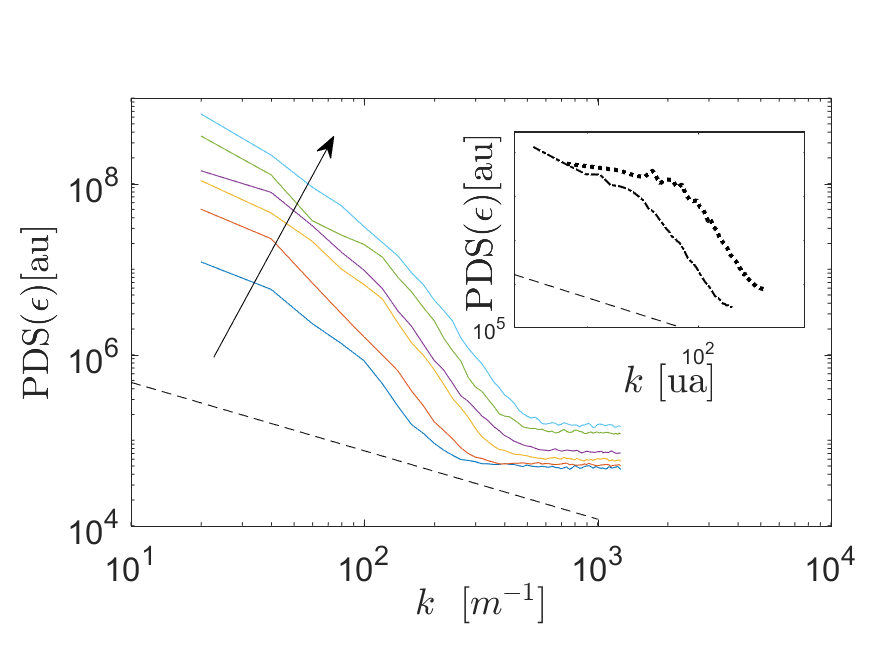}

\vspace{-0.2cm}

\caption{Power Density Spectra (PDS) of spatial fluctuations of $\epsilon$, averaged over time, as a function of wavenumber $k$, for $Re$ ranging from $2.4\times10^4$ to $1.68\times10^5$ (arrow indicates increasing $Re$). The inset shows the PDS of $\epsilon$ extracted from DNS at $Re=4\times 10^4$ in the spanwise (dotted line) and streamwise (dot-dashed line) directions.The dashed line in the main panel and the inset represents the power $k^{-(1-\mu)}$ predicted for the dissipation in HIT, with $\mu=0.2$.}
\label{PDSk}
\end{center}
\end{figure}

To explore further the similitudes and discrepancies of or measurements with the HIT reference, we performed a 2D Fourier transform of our dissipation maps. The 2D spectra were computed over the entire measurement window and averaged in time. The resulting Power Density Spectra (PDS) for $Re\in[0.24,1.68]\times10^5$ are shown in the main panel of figure \ref{PDSk}. We plot the PDS as a function of the radial wavenumber $k=\sqrt{k_x^2+k_y^2}$ because it is difficult to identify a privileged direction. As expected, the level of small-scale fluctuations increases with $Re$. Moreover, the spectra confirm that our pixel resolution is sufficient to capture all relevant structures, as we reach the noise level at high wavenumbers. This level is reached before $k_{\eta_K}=1/\eta_K$ that range from $3\times 10^3$ to $1.6\times 10^4~m^{-1}$, so we assume to be in the inertial range. $1.6\times 10^4~m^{-1}$, so we assume to be in the inertial range. However, the spectral decay is more pronounced than the power low in  $k^{-(1-\mu)}$ predicted for the HIT, as depicted by the dashed line in the main panel with $\mu=0.2$. The dissipation at the boundary does not behave like an imprint of the bulk where one expects to recover the universal behavior of the HIT in the inertial range.

To corroborate our results, we extracted similar quantities from the John Hopkins Turbulent Database. Although the energy injection differs between systems, we accessed near-wall dissipation from Direct Numerical Simulation (DNS) of a turbulent channel flow \cite{Li07,Graham16}. The simulation concerns a channel of size $8\pi\times3\pi\times2$ discretized over a grid of $2048\times1536\times512$ nodes. Periodic boundary conditions are applied in the x- and y-directions, and no-slip conditions in the z-direction. The Reynolds number is approximately $4\times10^4$. For more details about the DNS, see \cite{HopkinsChannel}. Our experimental observation window of  5.1$\times$5.1 cm$^2$ corresponds to approximately $20\times40$ nodes in the x- and y-directions and lies within the boundary layer. Dissipation maps from experiment and DNS (figures \ref{Snapshots}c and \ref{Snapshots}d, respectively) show similar structures, although anisotropy in the DNS is more pronounced due to the well-defined mean flow direction. More quantitatively, the PDF of fluctuations also matches a log-normal distribution (figure \ref{PDF}). Here, the decay of PDS in both directions is also much sharper than predicted for the HIT as shown in the inset of figure \ref{PDSk}. 

These numerical results reinforce our experimental findings, which can be summarized as follows. Using DWS, we were able for the first time to measure dissipative structures at the boundary of a turbulent flow, with good spatial and temporal resolution, provided the dynamics are sufficiently slow. Our measurements show that the boundary dissipation rate is proportional to the injected power and follows a pure power law scaling, without the logarithmic corrections typically observed in channel or pipe flows. At the impeller height, where measurements were made, there is an excess of dissipation that can be easily estimated. Dissipative structures fluctuate strongly and their fluctuations follow a log-normal distribution  as expected in HIT. In contrast, the PDS decay faster than expected in HIT. These two features are shared by  DNS of a channel flow. Therefore, we can hope that all properties measured in our experiment are representative of the dissipation at boundaries of a body impacted by a turbulent flow.\\

\noindent
{\bf Acknowledgements :}\\
The authors would like to thank B. Gallet, S. Fauve and M.E Brachet for fruitful discussions. This work is funded by the French National Research Agency (ANR DYSTURB Project No. ANR-17-CE30-0004).




\begin{thebibliography}{let1}
\bibitem{Francisco}
E. Francisco, V. Bouillaut, T. Wu, \& S. Aumaître,  Experiments in Fluids 64.9 (2023): 1
\bibitem{Li07}
Y. Li, E. Perlman, M. Wan, Y. Yang, C. Meneveau, R. Burns, S. Chen, A. Szalay \& G. Eyink. "A public turbulence database cluster and applications to study Lagrangian evolution of velocity increments in turbulence". Journal of Turbulence 9, No. 31, 2008. E. Perlman, R. Burns, Y. Li, and C. Meneveau. "Data Exploration of Turbulence Simulations using a Database Cluster". Supercomputing SC07, ACM, IEEE, 2007.
\bibitem{Graham16}
J. Graham, K. Kanov, X.I.A. Yang, M.K. Lee, N. Malaya, C.C. Lalescu, R. Burns, G. Eyink, A. Szalay, R.D.Moser.and C.Meneveau, Journal of Turbulence 17(2), 181 - 215, 2016. Taylor \& Francis ed.
\bibitem{K62}
A.N. Kolmogorov, J. Fluid Mech. {\bf 13}, 82 (1962)
\bibitem{Ahmed84}
S. R. Ahmed, G. Ramm and G. Faltin, 
 SAE Transactions, Vol. 93, Section 2: 840222––840402 (1984), pp. 473-503
 \bibitem{Lawn71}
J.C. Lawn, J . Fluid Mech. (1971), vol. 48, part 3, p p . 477-505
\bibitem{Fiedler00}
B. H. Fiedler, Q. J. R. Meteorol. SOC. (2000), 126, pp. 925-939
\bibitem{Asaro11}
E. D’Asaro, C. Lee, L. Rainville, R. Harcourt, L. Thomas, Sciences 332, (2011)
\bibitem{Yaglom}
A.S. Monin,  and A.M Yaglom {\it Statistical Fluid Mechanics, Mechanics of turbulence, } volume 2 (Edited by JL Lumley, The MIT press 1981)
\bibitem{Frish}
U. Frisch {\it Turbulence: The Legacy of A. N. Kolmogorov} (Cambridge University Press 1995)

\bibitem{Guichao2021}
G. Wang, F. Yang, K. Wu, Y. Ma, C. Peng, T. Liu and L-P. Wang, Chemical Engineering Science {\bf 229} (2021) 116133 , https://doi.org/10.1016/j.ces.2020.116133
\bibitem{Tsinober1992}
A. Tsinober, E Kit and T. Dracos (1992) J. Fluid Mech. {bf 242} p 169--192  
\bibitem{LathropNature}
B. W. Zeff, D. D. Lanterman, R. McAllister, R. Roy, E. J. Kostelich and  D. P. Lathrop, NATURE {\bf 421}--9 (2003)
\bibitem{Mullin}
J. A. Mullin and W. J. A. Dahm, Physics of Fluids {\bf 18}, 035102 (2006)
\bibitem{Schumacher}
J. Schumacher, arXiv:0710.4100v1 [physics.flu-dyn] 22 Oct 2007
\bibitem{Dosanjh85}
S. Dosanjh, J.A.C. Humphrey, Wear, Volume 102, Issue 4, 15 April 1985, Pages 309-330
\bibitem{Hogg97}
A.J. Hogg, H. E. Huppert and W. Brian Bade, J. Fluid Mech. (1997), vol. 338, pp. 317
\bibitem{Finnigan00}
J. Finnigan, {\it Turbulence in plant canopies} Annu. Rev. Fluid Mech. 2000. 32:519–571
\bibitem{Landau}
L.D. Landau, E.M. Lifshiftz {\it course in theoretical physics Vol. 6 Fluid Mechanics}
\bibitem{Pine88}
D. J. Pine, D. A. Weitz, P. M. Chaikin, and E. Herbolzheimer Phys. Rev. Lett. 60, 1134 (1988)
\bibitem{Pine91}
X-L. Wu {\it et al.}, In : J. Opt. Soc. Am. B 7.1 (1990), p. 15-20. doi : 10.1364/JOSAB.7.000015.
\bibitem{Bicout93}
D. Bicout  and R. Maynard, Physica A 199 (1993) 387-411
\bibitem{Bicout94}
D. Bicout, G. Maret, Physica A: Statistical mechanics and its applications, 210--1 (1994)
\bibitem{Durian91} 
D. J. Durian, D. A. Weitz et D. J. Pine, Phys. Rev. A 44--12 (1991), 
 \bibitem{Horn93}
 D.S. Horne et C.M. Davidson, In : Colloids and Surfaces
A : Physicochemical and Engineering Aspects 77.1 (1993). 
\bibitem{Menon97}
N. Menon et D.J. Durian, Science 275.5308
(1997), p. 1920-1922. 
\bibitem{Mason97}
T. G. Mason, Hu Gang et D. A. Weitz, J. Opt. Soc. Am. A
14.1 (1997), p. 139-149.
\bibitem{Palmer99}
A. Palmer {\it et al.}, Biophysical Journal 76.2 (1999), p. 1063-1071.
doi : https://doi.org/10.1016/S0006-3495(99)77271-1.
\bibitem{CohenAdda01} 
S. Cohen-Addad \& R. Höhler,
Phys. Rev. Lett. 86--20 (2001), p. 4700-4703
\bibitem{Crassous14}
A. Le Bouil, A. Amon, S. McNamara, and J. Crassous,
Phys. Rev. Lett. 112, 246001 (2014)
\bibitem{Przadka}
A. Przadka, B. Cabane, V. Pagneux,
A. Maurel, P. Petitjeans, Exp Fluids (2012) 52:519–527, DOI 10.1007/s00348-011-1240-x
\bibitem{Sheng} 
P. Sheng
{\it Introduction to Wave Scattering, Localization and Mesoscopic Phenomena} Second Edition (Springer Series in
materials science 2006)
\bibitem{Kienle}
A. Kienle {\it et al.}, Appl. Opt. 35.13 (1996), p. 2304-2314.
\bibitem{Faller21}
H. Faller {\it et al},  J. Fluid Mech. (2021), vol. 914, A2,
doi:10.1017/jfm.2020.908
\bibitem{Robinson91}
S.K. Robinson, Annu. Rev. Fluid Mech. 1991.23 : 601-.39
\bibitem{Sreenivasan}
K. R. Sreenivasan and P. Kailasnath, Physics of Fluids A: Fluid Dynamics, 5(2) (1993) 512-514.
\bibitem{Nelkin}
M. Nelkin,  Am. J. Phys. 68, 310–318 (2000)
\bibitem{Brachet}
M.E Brachet, Fluid Dynamics Research 8 (1991) l-8
\bibitem{HopkinsChannel}
Turbulent Channel Flow dataset: https://doi.org/10.7281/T10K26QW
\end{thebibliography}
\end{document}